\documentclass[conference,10pt]{IEEEtran}

\usepackage{amssymb}
\usepackage{amsmath}
\usepackage{amsthm}
\usepackage{amsfonts}
\usepackage{accents}
\usepackage{multirow}
\usepackage{array}
\usepackage{multirow}
\usepackage{graphicx}
\usepackage{epstopdf}
\usepackage{float}
\usepackage{pgfplots}
\pgfplotsset{compat=1.8}
\usepackage{epstopdf,epsfig}

\usepackage[switch]{lineno}
\usepackage{algorithm}

\ifCLASSOPTIONcompsoc
  \usepackage[caption=false,font=normalsize,labelfont=sf,textfont=sf]{subfig}
\else
  \usepackage[caption=false,font=footnotesize]{subfig}
\fi

\usepackage{balance}

\usepackage{booktabs}

\definecolor{mycolor1}{rgb}{0.00000,0.20000,0.60000}%
\definecolor{mycolor2}{rgb}{0.00000,0.40000,0.80000}%
\definecolor{mycolor3}{rgb}{0.40000,0.60000,1.00000}%
\definecolor{mycolor4}{rgb}{0.00000,0.60000,1.00000}%
\definecolor{mycolor5}{rgb}{1.00000,0.40000,0.60000}%
\definecolor{mycolor6}{rgb}{1.00000,0.20000,0.40000}%
\definecolor{mycolor7}{rgb}{1.00000,0.00000,0.20000}%
\definecolor{mycolor8}{rgb}{0.60000,0.20000,0.00000}%

\newtheoremstyle{colon}%
{}
{}
{\rm}
{}
{\itshape}
{:}
{ }
{\thmname{#1}\thmnumber{ \itshape#2}\thmnote{ (#3)}}

\theoremstyle{colon}
\begin{document}
\title{Study of Tomlinson-Harashima Precoders for Rate-Splitting-Based Cell-Free MIMO Networks}

\author{André R. Flores$^1$ and Rodrigo C. de Lamare$^{1,2}$ \\
$^1$Centre for Telecommunications Studies, Pontifical Catholic
University of Rio de Janeiro, Brazil \\
$^2$Department of Electronic Engineering, University of York,
United Kingdom \\
Emails: andre\_flores@esp.puc-rio.br, delamare@puc-rio.br }

\maketitle

\begin{abstract}
Cell-free (CF) systems have the potential to fulfill the increasing performance demand of future wireless applications by employing distributed access points (APs) that transmit the information over the same time-frequency resources. Due to the simultaneous transmission,  multiuser interference (MUI) degrades the overall performance. To cope with the MUI in the downlink several linear precoding techniques, which rely on perfect channel state information at the transmitter (CSIT), have been studied. However, perfect CSIT is hardly obtained in practical systems. In this context, rate-splitting (RS) has arisen as a potential solution to deal with CSIT imperfections. In contrast to existing works, we explore non-linear precoding techniques along with RS-CF systems. Furthermore, the multi-branch (MB) concept is included to further enhance the overall performance of the system. Simulations show that the proposed MB-THP for RS-based CF systems outperforms the conventional linear precoders. 
\end{abstract}


%

\section{Introduction}
Future applications demand new wireless communications architectures, capable of offering higher throughput than the coordinated and centralized base stations (BSs) currently deployed \cite{Tataria2021,mmimo,wence}. Cell-free (CF) multiple-input multiple-output (MIMO) systems have have the potential to improve the overall performance and fulfill future applications requirements \cite{Ammar2022}. In contrast to BS-based networks, CF-MIMO  deploys multiple access points (APs) to cover the geographical area of interest. A central processing unit (CPU), which may be located at the cloud server, controls the APs. CF deployment outperforms BSs systems in terms of throughput per user \cite{Elhoushy2021b} and energy efficiency \cite{Jin2021}. 

To cope with the multiuser interferecen (MUI) in the downlink, CF MIMO systems employ precoding techniques. Linear precoding such as matched filter (MF), zero-forcing (ZF) \cite{Nayebi2017,siprec,lcbd,wlbd,bbprec}, and minimum mean-square error (MMSE) \cite{rmmseprec,rmmse,Bjoernson2020,zcprec} precoders have been reported in the literature. Initially, these techniques were employed as network-wide (NW) precoders \cite{Ngo2017}, which are not practical because of the extremely high signaling load and computational cost. 

Clusters of APs and users can be employed to reduce the computational complexity and the signaling load \cite{Bjoernson2020a,Palhares2020,Flores2023,esg,cl&sched}. In \cite{Palhares2020}, the number of APs is curtailed to make better use of the resources. Scalable MMSE 
combiners are derived along with precoders by exploiting the uplink-downlink duality in \cite{Bjoernson2020a}. A regularized ZF precoder is proposed in \cite{Lozano2021}, where several subsets are formed to reduce the number of APs serving each user. In \cite{Flores2022}, clusters of users and APs are formed to reduce the overall computational complexity. 

One major problem in the design of CF precoders is to assume perfect knowledge of the channel state information at the transmitter (CSIT) . In practice, such assumption is hardly verified leading to residual multiuser interference (MUI) which is detrimental for the overall performance of the sytem \cite{Vu2007}. As a result, rate splitting (RS) has emerged as 
a potential technology to deal better with imperfect CSIT. This non-conventional transmit scheme  enhances the robustness of the system by splitting the data into
common messages and private messages \cite{Joudeh2016,Mao2022}. Both messages are superimposed before the transmission. The main advantage is that RS provides flexibility to the system since it can adapt the content and the power of the common stream. In fact, it has been reported in the literature that RS outperforms conventional transmit schemes such as spatial division multiple access (SDMA) and non-orthogonal multiple access (NOMA)\cite{Clerckx2020}. Due to its benefits, RS has been studied with linear \cite{Joudeh2016} and non-linear \cite{Flores2021} precoders and CF systems\cite{Flores2022a,Mishra2022}.

In contrast to previous works, we propose a Tomlinson-Harashima precoder (THP) \cite{mbthp,rmbthp,rsthp} for RS-based CF systems working with clusters of APs. The proposed cluster-based THP keeps the signaling load low while enhancing the overall performance. Moreover, we incorporate the multi-branch (MB) approach, which optimizes user ordering to obtain even a better performance. Numerical experiments show that the proposed nonlinear techniques outperforms existing precoders. 

The rest of this paper is organized as follows. In section II, we present the conventional system model of a CF system and describe the RS and THP transmission schemes. In section III, the proposed THP for RS-based CF system is derived. The ergodic sum-rate 
(ESR), which is the metric employed to evaluate the performance of the proposed technique, is introduced in section IV. Simulation results are depicted in section V. Finally, Section VI concludes this paper.

We employ bold lowercase, bold uppercase, and calligraphic letters for the vectors, matrices, and sets respectively; The trace and statistical expectation operators are denoted by $\textrm{Tr}(\cdot)$ and $\mathbb{E}\left[\cdot\right]$, respectively; the notations $(\cdot)^{\text{T}}$, $(\cdot)^H$, and $\lVert\cdot\rVert$  stand for the transpose, Hermitian, and Euclidean norm, respectively. The real part of a complex input 
 is obtained with the operator $\text{Re}\left( \cdot \right)$ whereas $\text{Im}\left( \cdot \right)$ denotes the imaginary part.

\section{System Model}

Let us consider the downlink of a CF system operating with $M$ single-antenna APs, which are distributed over the area of interest. The system provides service to $K$ single-antenna users, which are spread out over the area of interest. The system operates in an underloaded regime, i.e., $M>K$. The transmission takes place over a fading channel given by $\mathbf{G}=\left[\mathbf{g}_1,\mathbf{g}_2,\cdots,\mathbf{g}_K\right]\in\mathbb{C}^{M\times K}$, where $\mathbf{g}_k$ contains the channel connecting each AP to the $k$-th user. Following this model, the received signal at user $k$ is given by
\begin{equation}
    y_k=\mathbf{g}_k^{H}\mathbf{x}+n_k,
\end{equation}
where $\mathbf{x}\in\mathbb{C}^{M}$ is the transmitted vector and $n_k$ is a complex random variable that represents the additive white Gaussian noise (AWGN) with zero mean and unit variance, i.e., $n_k\sim\mathcal{CN}\left(0,\sigma_n^2\right)$. Moreover, the system follows a transmit power constraint, i.e., $\mathbb{E}\left[\lVert\mathbf{x}\rVert^2\right]<P_t$, where $P_t$ denotes the available power for transmission.

The system employs the time division duplexing (TDD) protocol, which allow us to obtain the channel estimate $\hat{\mathbf{G}}^{H}\in\mathbb{C}^{K\times M}$ by employing pilot training and the reciprocity property \cite{Vu2007}. Specifically, the central processing unit (CPU) computes $\hat{\mathbf{G}}^{H}=\left[\hat{\mathbf{g}}_1,\hat{\mathbf{g}}_2,\cdots,\hat{\mathbf{g}}_K\right]^{H}$. The $m$-th coefficient of vector $\hat{\mathbf{g}}_k$ represents the channel estimate that links the $m$-th AP with user $k$ and  is given by
\begin{equation}
    \hat{g}_{m,k}=\sqrt{\zeta_{m,k}}\left(\sqrt{1-\sigma_e^2}h_{m,k}+\sigma_e\tilde{h}_{m,k}\right),
\end{equation}
where $\zeta_{m,k}$ denotes the large-scale fading coefficient, which incorporates the path-loss and shadowing effects; the coefficients $h_{m,k}$ are independent and identically distributed (i.i.d.) random variables that represent the small-scale fading and follow a complex Gaussian distribution $\mathcal{CN}\left(0,1\right)$; the coefficients $\tilde{h}_{m,k}$ are i.i.d random variables, also independent from $h_{m,k}$, that follow a complex Gaussian distribution $\mathcal{CN}\left(0,1\right)$ and model the imperfections in the channel estimate; $\sigma_e$ represents the CSIT quality. Then, the error affecting the coefficients $\hat{g}_{m,k}$ is given by $\tilde{g}_{m,k}=\sigma_e\sqrt{\zeta_{m,k}}\tilde{h}_{m,k}$.

\subsection {Rate-Splitting}

Let us consider an RS scheme where the message of one user is split into a common part and a private part. The common part is encoded into a common stream. The private part and the messages of the other users are encoded into $K$ private streams. It follows that each user must decode two different streams. Specifically, the common stream is decoded first by all users and the private streams are treated as noise. Note that although the common stream is decoded by all users, it is only intended to one user. Once the common stream is decoded, the system employs successive interference cancellation (SIC) \cite{itic,jidf,spa,mfsic,dfcc,vfap,did,mbdf,bfidd,1bitidd,comp,list_mtc,det_mtc,msgamp,msgamp2,cfidd,llrref} to remove the common stream from the received signal. Finally, each user decodes its private stream by treating the other private streams as noise. 

By splitting one message, $K+1$ streams are originated. Then, the vector of data symbols is given by $\mathbf{s}^{\left(\text{RS}\right)}=\left[s_c,s_1,s_2,\cdots,s_K\right]^{\text{T}}$, where $s_c$ denotes the symbol corresponding to the common stream and $s_k$ denotes the private symbol of user $k$. A precoder $\mathbf{P}^{\left(\text{RS}\right)}=\left[\mathbf{p}_c,\mathbf{p}_1,\mathbf{p}_2,\cdots,\mathbf{p}_K\right]\in \mathbb{C}^{M\times \left(K+1\right)}$ maps the symbols to the APs. In particular, the common precoder $\mathbf{p}_c$ maps the common symbol to the APs, whereas the private precoder $\mathbf{p}_k$ maps the symbol intended to the $k$-th user to the APS. It follows that the transmitted signal is given by
\begin{equation}
    \mathbf{x}^{\left(\text{RS}\right)}=s_c\mathbf{p}_c+\sum_{k=1}^{K}s_k\mathbf{p}_k.
\end{equation}

The total power is allocated partially to the common stream and partially to the private streams. In fact, several resource allocation techniques \cite{tds,jpba} can be adapted or directly employed for power adjustment and AP selection. However, the system must obey the transmit power constraint and therefore we have
\begin{equation}
    \mathbb{E}\left[\lVert\mathbf{x}\rVert^2\right]=\lVert\mathbf{p}_c\rVert^2+\sum_{k=1}^{K} \lVert\mathbf{p}_k\rVert^2\leq P_t.
\end{equation}
Thus, $\lVert\mathbf{p}_c\rVert^2=0$ means that no power is allocated to the common stream and no RS is preformed. In other words, the system is simplified to the conventional CF system.

\subsection{Tomlinson-Harashima Precoder}
THP was originally designed to deal with channels with intersymbol interference \cite{Tomlinson1971,Harashima1972}. Due to its benefits, it was further extended to MIMO systems employing spatial division multiple access (SDMA) in \cite{Fischer2002}. Interestingly, THP can be interpreted as the counterpart of SIC implemented at the receiver \cite{Windpassinger2004} and therefore the ordering of the precoded symbols impacts directly in the performance  of the system. 

The conventional THP algorithm implements three filters, namely a feedback filter $\mathbf{B}\in\mathbb{C}^{K\times K}$ that deals with the MUI by suppressing the interference of the previous symbols, a scaling matrix $\mathbf{C}\in\mathbb{C}^{K\times K}$ that weights each stream of data and a feedforward filter $\mathbf{F}\in\mathbb{C}^{M\times K}$ that enhances spatial causality. Since matrix $\mathbf{B}$ deal with the interference of previous symbols, it must have a lower triangular structure. On the other hand, matrix $\mathbf{C}$ is a diagonal matrix.  

Two general THP structures have been reported in literature. The difference between these to structures is the position of the scaling matrix $\mathbf{C}$. The centralized THP (cTHP) implements the scaling matrix at the transmitter side. In contrast, the decentralized THP (dTHP) considers that the scaling matrix is located at the receiver side. 

The three filters employed on the THP algorithm are defined based on the specific THP structure that is being implemented. In particular, the zero-forcing THP (ZF-THP), which aims to remove the MUI completely, can be implemented by performing an LQ decomposition given by $\hat{\mathbf{G}}^{H}=\mathbf{L}\mathbf{Q}$, where $\mathbf{L}\in\mathbb{C}^{K\times K}$ is a lower triangular matrix and $\mathbf{Q}\in\mathbb{C}^{K\times M}$ is a unitary matrix. Then, we have
\begin{align}
\mathbf{F}&={\mathbf{Q}}^{H},\\
\mathbf{C}&=\text{diag}\left({l}_{1,1},{l}_{2,2},\cdots,{l}_{K,K}\right),\\
\mathbf{B}^{\left(\text{c}\right)}&={\mathbf{L}}\mathbf{C}, \\\quad\mathbf{B}^{\left(\text{d}\right)}&=\mathbf{C}{\mathbf{L}},
\end{align}
where $l_{k,k}$ denote the $k-$th diagonal coefficient of matrix $\mathbf{L}$, and $\left(\cdot\right)^{\left(\text{c}\right)}$, $\left(\cdot\right)^{\left(\text{d}\right)}$ denote that the filter is used in the centralized and decentralized structure, respectively.

Note that the feedback processing increases the amplitude of the transmitted symbols thereby introducing a power loss so that the power constraint is fulfilled. A modulo operation is included in the THP algorithm to mitigate the power loss and reduce the amplitude of the transmitted symbols. The modulo operation is given by
\begin{equation}
    \mathcal{M}\left(\Breve{s}_k\right)=\Breve{s}_k-\left\lfloor\frac{\text{Re}\left(\Breve{s}_k\right)}{\lambda}+\frac{1}{2}\right\rfloor\lambda-j\left\lfloor\frac{\text{Im}\left(\Breve{s}_k\right)}{\lambda}+\frac{1}{2}\right\rfloor\lambda,
\end{equation}
where $\lambda$ depends on the power of the transmitted symbol and the modulation alphabet. A typical value for QPSK and symbols with unit variance is $\lambda=2\sqrt{2}$.

\section{Proposed THP for RS-based CF systems}
To further enhance the performance of CF systems, a RS-based transmission scheme with THP is considered. In addition, AP selection is performed to reduce the signaling load. In particular, the proposed scheme employs nonlinear precoders to transmit the data streams to the distributed APs. Specifically, the ZF-THP is employed to precode the private streams. On the other hand, the common message is linearly precoded. 

Let us first reduce the signaling load by taking into account that only a small cluster of APs transmit the most relevant part of the signal. Other APs may be discarded since its contribution to the transmit signal is small. To select the APs, we consider the ones with the largest large-scale fading coefficient and then we gather the APs in the set $\mathcal{A}_k$. Therefore the number of APs serving the $k$-th user is given by $\lvert\mathcal{A}_k\rvert$.  AP selection results in an equivalent channel estimate given by $\bar{\mathbf{G}}^{H}=\left[\mathbf{\bar{g}}_1,\mathbf{\bar{g}}_2,\cdots,\mathbf{\bar{g}}_k\right]^{H}\in \mathbb{C}^{K \times M}$, which is a sparse matrix. The $m$-th coefficient of vector $\bar{\mathbf{g}}_k$ is given by
\begin{equation}
    \bar{g}_{m,k}=\begin{cases}
\hat{g}_{m,k},&m\in \mathcal{A}_k,\\
0, &\text{otherwise.}
\end{cases}
\label{sparse effective channel}
\end{equation}

The performance of the THP algorithm depends directly on the symbol ordering. Therefore, we employ a technique known as the MB approach, where multiple candidate branches are generated. For instance, let us consider that $L$ ordering patterns are generated. Let us now introduce matrix $\mathbf{T}_l$, $l=1,2,\cdots,L$, which represents the $l$-th transmit pattern. Once the patterns are defined, the system chooses the branch leading to the highest performance.

The patterns are designed and pre-stored according to the following equation: 
\begin{align}
    \mathbf{T}_1&=\mathbf{I}_{K},\nonumber\\
    \mathbf{T}_l&=\begin{bmatrix}
        &\mathbf{I}_q     &\mathbf{0}_{q,M-q}\\
        &\mathbf{0}_{M-q,q} &\mathbf{\Pi}_{M-q}
    \end{bmatrix},
\end{align}
where $q=l-2$. The matrix $\mathbf{\Pi}\in\mathbb{R}^{\left(M-q\right)\times\left(M-q\right)}$ exchange the order of the users, thereby exchanging the order of the transmitted symbols. The coefficients of $\mathbf{\Pi}$ are zero except on the reverse diagonal where they take the value of one. 

After selecting the best branch among the patterns, the equivalent channel is rearranged to take into account the symbol order scheme, i.e., $\bar{\mathbf{G}}_o^{H}=\mathbf{T}_o\bar{\mathbf{G}}^H$, where $\mathbf{T}_o$ denotes the selected transmission pattern. Once the channel matrix is defined, we can proceed to the design of the precoder. For this purpose, we have to take into account that the feedback processing can be implemented with a matrix inversion and the modulo operation is identical to add a perturbation vector $\mathbf{d}\in\mathbb{C}^{K}$ to the transmitted symbols, i.e., $\check{\mathbf{s}}=\mathbf{s}+\mathbf{d}$. Then, the transmitted signal can be expressed as $\mathbf{x}_{o}^{\left(\text{RS}\right)}=\left[\mathbf{p}_{c,o},\mathbf{P}_{p,o}\right]\left[s_c,\check{\mathbf{s}}_{o}\right]$, where the subscript $\left(\cdot\right)_o$ denotes the use of the best branch and $\mathbf{P}_{p,o}\in \mathbb{C}^{M\times K}$ is the nonlinear private precoder computed using the best branch. Specifically, we employ an LQ decomposition given by $\bar{\mathbf{G}}_o^{H}=\bar{\mathbf{L}}_o\bar{\mathbf{Q}}_o$ to get the private precoder $\mathbf{P}_{p,o}$. It follows that \begin{align}
\bar{\mathbf{F}}_o&={\bar{\mathbf{Q}}}_o^{H},\\
\bar{\mathbf{C}}_o&=\text{diag}\left({\bar{l}}_{1,1},{\bar{l}}_{2,2},\cdots,{\bar{l}}_{K,K}\right),\\
\bar{\mathbf{B}}_o^{\left(\text{c}\right)}&=\bar{\mathbf{L}}_o\bar{\mathbf{C}}_o, \\\quad\mathbf{B}_o^{\left(\text{d}\right)}&=\bar{\mathbf{C}}_o{\bar{\mathbf{L}}_o}.
\end{align}

Depending on the THP structure employed, we obtain
\begin{align}
    \mathbf{P}_{p,o}^{\left(c\right)}&=\beta_o^{\left(\text{c}\right)}\bar{\mathbf{F}}_o\bar{\mathbf{C}}_o\bar{\mathbf{B}}_o^{\left(\text{c}\right)^{-1}}, \\
    \mathbf{P}_{p,o}^{\left(d\right)}&=\beta_o^{\left(\text{d}\right)}\bar{\mathbf{F}}_o\bar{\mathbf{B}}_o^{\left(\text{d}\right)^{-1}},
\end{align}
where $\beta^{\left(\text{c}\right)}_o$ and $\beta^{\left(\text{d}\right)}_o$ denote the power scaling factor introduced to satisfy the transmit power constraint. Note that the power allocated to the common stream is equal to $\lVert\mathbf{p}_{c,o}\rVert^2$. Then, the power scaling factor is equal to 
\begin{align}
    \beta_o^{\left(\text{c}\right)}=\sqrt{\frac{P_t-\lVert\mathbf{p}_{c,o}\rVert^2}{\sum_{k=1}^K l_{k,k}}},\\
    \beta_o^{\left(\text{d}\right)}=\sqrt{\frac{P_t-\lVert\mathbf{p}_{c,o}\rVert^2}{K}}.
\end{align}

The signal arriving at the receivers can be expressed as follows:
\begin{align}
    \mathbf{y}_o^{\left(\text{c}\right)}&=\frac{1}{\beta_o^{\left(c\right)}}\mathbf{G}_o^{H}\mathbf{p}_{c,o} s_c+\check{\mathbf{s}}_o+\frac{1}{\beta^{\left(c\right)}}\mathbf{n},\\
    \mathbf{y}_o^{\left(\text{d}\right)}&=\frac{1}{\beta_o^{\left(d\right)}}\bar{\mathbf{C}}_o\mathbf{G}_o^{H}\mathbf{p}_{c,o} s_c+\check{\mathbf{s}}_o+\frac{1}{\beta^{\left(d\right)}}\bar{\mathbf{C}}_o\mathbf{n}.
\end{align}
Specifically, at the $k$-th receiver, we obtain
\begin{align}
    y_{k,o}^{\left(\text{c}\right)}&=\frac{1}{\beta_{o}^{\left(c\right)}}\mathbf{g}_{o}^{H}\mathbf{p}_{c,o} s_c+\check{s}_k+\frac{1}{\beta_{o}^{\left(c\right)}}n_k,\label{cTHP received signal at user k }\\
    y_{k,o}^{\left(\text{d}\right)}&=\frac{1}{\beta_{o}^{\left(d\right)}l_{k,k}}\mathbf{g}_{o}^{H}\mathbf{p}_{c,o} s_c+\check{s}_k+\frac{1}{\beta_{o}^{\left(d\right)}l_{k,k,o}}n_k.\label{dTHP received signal at user k }
\end{align}
From \eqref{cTHP received signal at user k } and \label{dTHP received signal at user k } we can compute the average power of the receive signal, i.e., $\mathbb{E}\left[\lVert y_k\rVert^2\right]$. Then, we arrive to the following signal-to-interference-plus-noise (SINR) ratios:
\begin{align}       
    \gamma_{c,k,o}^{\left(\text{c}\right)}&=\frac{\sum\limits_{j=1}^K\frac{1}{l_{j,j}}\lvert\mathbf{g}_{k,o}^{H}\mathbf{p}_{c,o}\rvert^2}{P_t-\lVert\mathbf{p}_{c,o}\rVert^2-\sigma_n^2\sum\limits_{j=1}^K\frac{1}{l_{j,j,o}}}\\
    \gamma_{c,k,o}^{\left(\text{d}\right)}&=\frac{K\lvert\mathbf{g}_{k,o}^{H}\mathbf{p}_{c,o}\rvert^2}{l_{k,k,o}^2\left(P_t-\lVert\mathbf{p}_{c,o}\rVert^2\right)+K\sigma_n^2}
\end{align}

Once the common signal is removed, the SINR when decoding the private streams is given by
\begin{align}       
    \gamma_{k,o}^{\left(\text{c}\right)}&=\frac{P_t-\lVert\mathbf{p}_{c,o}\rVert^2}{\sum\limits_{j=1}^K\frac{1}{l_{j,j,o}}}\\
    \gamma_{k,o}^{\left(\text{d}\right)}&=\frac{l_{k,k}^2\left(P_t-\lVert\mathbf{p}_{c,o}\rVert^2\right)}{K\sigma_n^2}
\end{align}

\section{Ergodic Sum-Rate}
Since imperfect CSIT is considered, the instantaneous rates given a channel realization are not achievable. Therefore, we adopt the ESR as the main metric to evaluate the performance of the proposed system. The ESR is defined as 
\begin{equation}
    S_e^{\left(\text{RS}\right)}=\min_{k\in [1,K]}\mathbb{E}\left[\bar{R}_{c,k}\right]+\mathbb{E}\left[\bar{R}_p\right],
\end{equation}
where $\bar{R}_{c,k}$ represents the average common rate of the $k$-th user and $\bar{R}_p$ stands for the average sum-private rate of the system. Note that we set the ergodic common rate to the minimum found across all users to guarantee that all users decode the common message. 

The average sum-private rate is equal to $\bar{R}_p=\mathbb{E}\left[R_p|\hat{\mathbf{G}}^{H}\right]$, where $R_p$ stands for the instantaneous sum-private rate. $R_p$ can be computed as the sum of all the instantaneous rates of the users, i.e., $R_p=\sum_{k=1}^K R_k$, where $R_k$ is the instantaneous private rate of user $k$. Similarly, the average common rate at user $k$ can be obtained as the expected value of the instantaneous common rate given a channel realization, i.e., $\bar{R}_{c,k}=\mathbb{E}\left[R_{c,k}|\bar{\mathbf{G}}^{H}\right]$, where $R_{c,k}$ denotes the instantaneous common rate at user $k$. 

The average rates not only allow us to compute the ESR, but it is also the criterion adopted to select the best transmission pattern. Specifically, the system selects the branch that maximizes the average sum-private rate, i.e., 
\begin{equation}
    \mathbf{T}_o=\max_l\mathbb{E}=\left[\bar{R}_p|\bar{\mathbf{G}}_l^{H}\right],
\end{equation}
where $\bar{\mathbf{G}}_l^{H}=\mathbf{T}_l\bar{\mathbf{G}}^{H}$ denotes the channel rearranged according to the $l$-th branch.

Assuming Gaussian codebooks, the instantaneous common rate is defined by
\begin{equation}
    R_{c,k}=\log_2\left(1+\gamma_{c,k}\right),
\end{equation}
where $\gamma_{c,k}$ is the signal-to-interference-plus-noise ratio (SINR) at user $k$ when decoding the common stream. On the other hand, we have 
\begin{equation}
    R_{k}=\log_2\left(1+\gamma_k\right),
\end{equation}
where $\gamma_k$ is the SINR when decoding the private stream at user $k$.

\section{Numerical Results}
We assess the performance of the proposed TH precoders via numerical experiments. Throughout the experiments, the large scale fading coefficients are set to
\begin{equation}
     \zeta_{k,n}=P_{k,n}\cdot 10^{\frac{\sigma^{\left(\textrm{s}\right)}z_{k,n}}{10}},
 \end{equation}
 where $P_{k,n}$ is the path loss and the scalar $10^{\frac{\sigma^{\left(\textrm{s}\right)}z_{k,n}}{10}}$ include the shadowing effect with standard deviation $\sigma^{\left(\textrm{s}\right)}=8$. The random variable $z_{k,n}$ follows Gaussian distribution with zero mean and unit variance. The path loss was calculated using a three-slope model as\par\noindent\small
 \begin{align}
     P_{k,n}=\begin{cases}
  -L-35\log_{10}\left(d_{k,n}\right), & \text{$d_{k,n}>d_1$} \\
  -L-15\log_{10}\left(d_1\right)-20\log_{10}\left(d_{k,n}\right), & \text{$d_0< d_{k,n}\leq d_1$}\\
    -L-15\log_{10}\left(d_1\right)-20\log_{10}\left(d_0\right), & \text{otherwise,}
\end{cases}
 \end{align}\normalsize
 where $d_{k,n}$ is the distance between the $n$-th AP and the $k$-th user, $d_1=50$ m, $d_0= 10$ m, and the attenuation $L$ is \par\noindent 
 \begin{align}
 L=&46.3+33.9\log_{10}\left(f\right)-13.82\log_{10}\left(h_{\textrm{AP}}\right)\nonumber\\
     &-\left(1.1\log_{10}\left(f\right)-0.7\right)h_u+\left(1.56\log_{10}\left(f\right)-0.8\right),
 \end{align}\normalsize
 where $h_{\textrm{AP}}=15$ m and $h_{u}=1.65$ m are the positions of the APs and UEs above the ground, respectively. We consider a frequency of $f= 1900$ MHz. The noise variance is 
  \begin{equation}
     \sigma_n^2=T_o k_B B N_f,
 \end{equation}
 where $T_o=290$ K is the noise temperature, $k_B=1.381\times 10^{-23}$ J/K is the Boltzmann constant, $B=50$ MHz is the bandwidth and $N_f=10$ dB is the noise figure. The signal-to-noise ratio (SNR) is \par\noindent
 \begin{equation}
     \text{SNR}=\frac{P_{t}\textrm{Tr}\left(\mathbf{G}^{\text{H}}\mathbf{G}\right)}{N K \sigma_n^2},
 \end{equation}\normalsize
where $\textrm{Tr}(\cdot)$ is the trace of its matrix argument.

For all experiments, we have $12$ APs randomly distributed over a square with side equal to $400$ m. The APs serve a total of $3$ users, which are geographically distributed. We considered a total of 10,000 channel realizations to compute the ESR. Specifically, we employed $100$ channel estimates and, for each channel estimate, we considered $100$ error matrices. It follows that the average rate was computed with $100$ error matrices. We employ an SVD over the equivalent channel, i.e. $\bar{\mathbf{G}}_o=\bar{\mathbf{U}}_o\bar{\mathbf{\Psi}}_o\bar{\mathbf{V}}_o$. Then, we set the common precoder equal to the first column of matrix $\bar{\mathbf{V}}_o$, i.e., $\mathbf{p}_{c,o}=\bar{\mathbf{v}}_{1,o}$. 

\begin{figure}[t]
\begin{center}
\includegraphics[width=1\columnwidth]{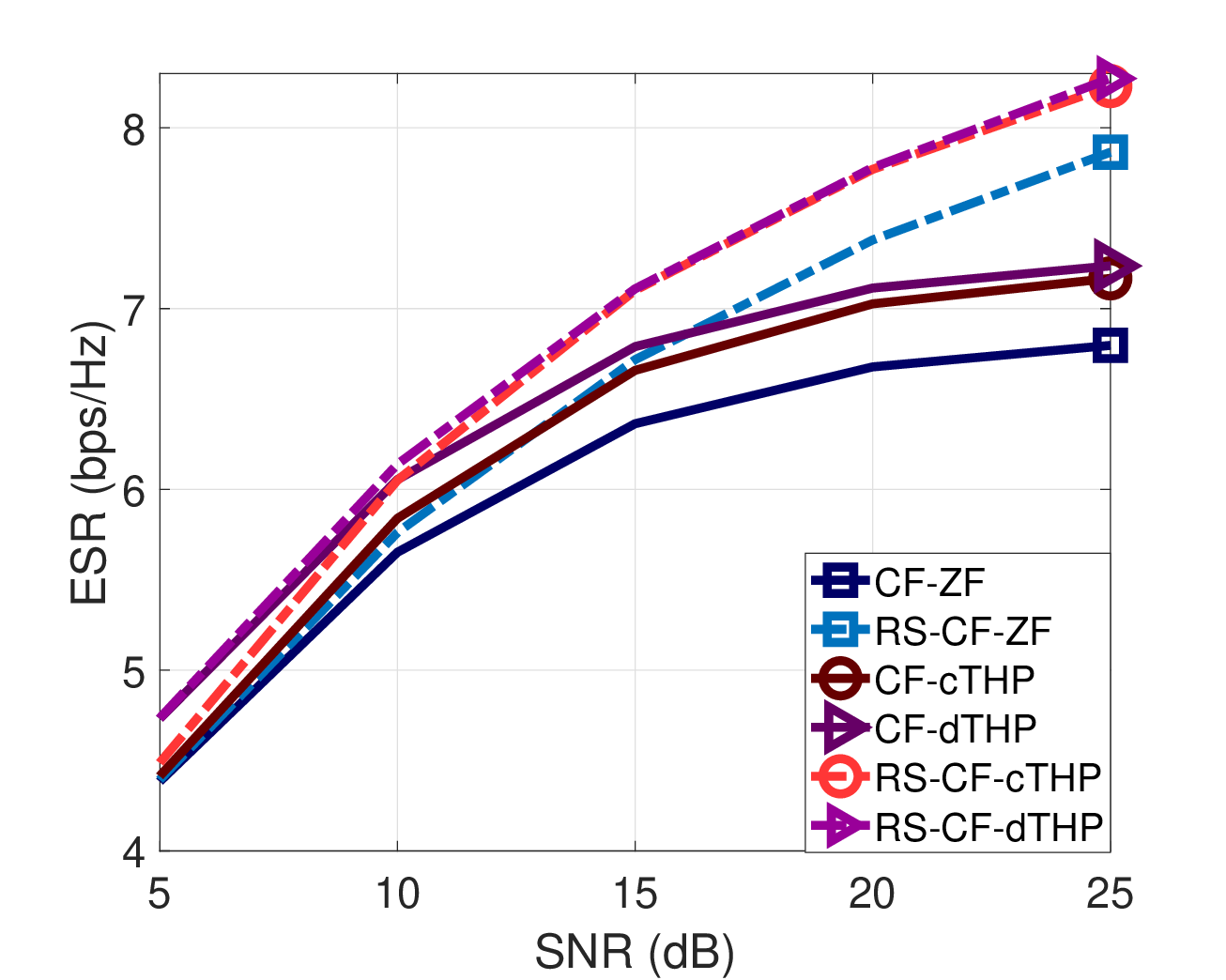}
\vspace{0.1em}
\caption{Sum-rate performance of linear and nonlinear ZF precoders versus SNR. Here, $M=12$, $K=3$, $|\mathcal{A}_k|=6$, $\sigma_{e}^2=0.15$. }
\label{Fig1}
\end{center}
\end{figure}

Fig. \ref{Fig1} depicts the ESR obtained by different ZF precoding techniques under imperfect CSIT. The proposed THP for RS-based CF systems outperforms the conventional ZF linear precoder, enhancing the capabilities of the CF network. The gains obtained by the proposed techniques increases with the SNR, meaning that this technique deals better with CSIT imperfections. Moreover, the decentralized structure achieves a slightly better performance than the centralized one at the cost of increasing slightly the receive processing operations.

\begin{figure}[t]
\begin{center}
\includegraphics[width=1\columnwidth]{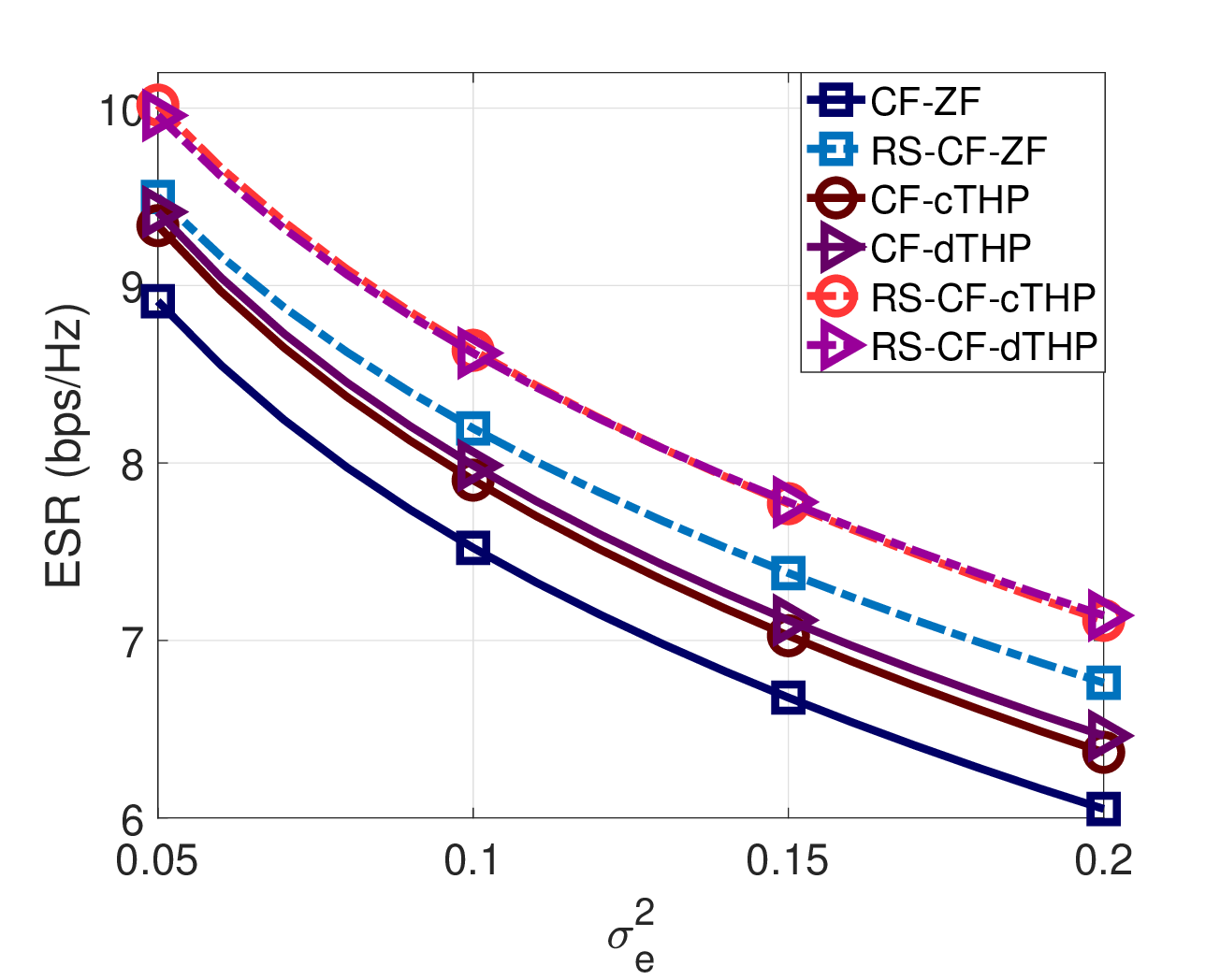}
\vspace{0.1em}
\caption{Sum-rate performance of linear and nonlinear ZF precoders versus $\sigma_e^2$. Here, $M=12$, $K=3$, $|\mathcal{A}_k|=6$, $\text{SNR}=20$ dB. }
\label{Fig2}
\end{center}
\end{figure}

Fig. \ref{Fig2} shows the ESR performance against different CSIT qualities. The CSIT imperfections directly affect the overall performance and therefore all curves show a decreasing trend. The best performance is obtained by the proposed cTHP and dTHP for RS-based CF systems. This result shows the effectiveness of our approach against CSIT imperfections, which is one of the major problems in CF systems. 

\begin{figure}[t]
\begin{center}
\includegraphics[width=1\columnwidth]{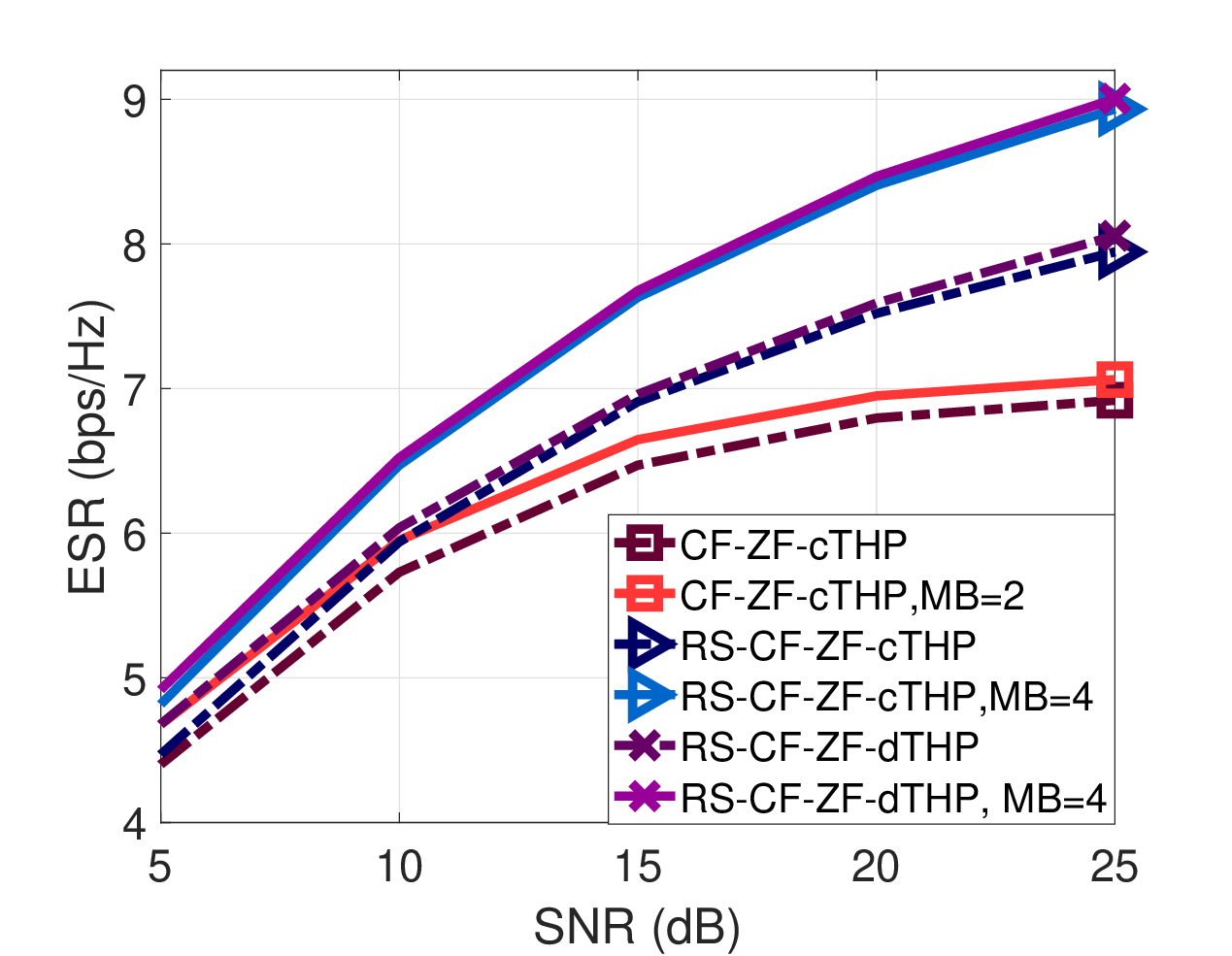}
\vspace{0.1em}
\caption{Sum-rate performance of multi-branch nonlinear precoders versus SNR. Here, $M=12$, $K=3$, $|\mathcal{A}_k|=6$, $\sigma_{e}^2=0.15$. }
\label{Fig3}
\end{center}
\end{figure}

In the last example, we assess the performance of the proposed multi-branch technique RS-based CF systems precoded with THP. Figure \ref{Fig3} shows that the multi-branch approach increases the ESR performance for all THP. The best performance is obtained by the ZF-dTHP with four branches for RS-based CF systems. We can also conclude that the gain increases as the number of branches increases.  

\section{Conclusions}

We proposed nonlinear precoders for RS-based CF systems, which are based on the noninear THP scheme. The proposed precoders can employ two different THP structures, namely, cTHP and dTHP structure. AP selection was performed to reduce the signaling load, which produces sparse channel matrices. Simulations showed that the proposed RS-based cTHP and dTHP for CF achieve better ESR than conventional linear approaches. Morover, simulation results confirmed the effectiveness of our approach against different levels of CSIT uncertainties when compared to conventional schemes.

\bibliographystyle{IEEEtran}
\bibliography{SubsetBib2}
\end{document}